\documentclass[conference]{IEEEtran}
\IEEEoverridecommandlockouts
\usepackage{cite}
\usepackage{amsmath,amssymb,amsfonts}
\usepackage{algorithmic}
\usepackage{graphicx}
\usepackage{textcomp}
\usepackage{xcolor}

\def\BibTeX{{\rm B\kern-.05em{\sc i\kern-.025em b}\kern-.08em
    T\kern-.1667em\lower.7ex\hbox{E}\kern-.125emX}}

\begin{document}

\title{Vibe2Spike: Batteryless Wireless Tags for Vibration Sensing with Event Cameras and Spiking Networks}

\author{
\IEEEauthorblockN {
Danny~Scott$^{*}$, 
William~LaForest$^{*}$, 
Hritom~Das$^{\$}$,   
Ioannis~Polykretis$^{\dag}$,  \\
Catherine~D.~Schuman$^{*}$,
Charles~Rizzo$^{*}$,
James~Plank$^{*}$,
and~Sai~Swaminathan$^{*}$ 
}
\IEEEauthorblockA {
$^{*}$Department of EECS, The University of Tennessee, Knoxville, Knoxville, TN 37996 USA \\
$^{\$}$Department of ECE, Oklahoma State University, OK 74078, USA \\
$^{\dag}$Accenture Labs, Accenture, San Francisco CA, 94105 USA. 
}
\IEEEauthorblockA {
$^{*}${dscott57, wlafore2, crizzo, cschuman, jplank, sai}@utk.edu; 
$^{\$}$hritom.das@okstate.edu;
$^{\dag}$ioannis.polykretis@accenture.com;
}

}

\maketitle

\begin{figure*}[!ht]
  \centering
  \includegraphics[width=\textwidth]{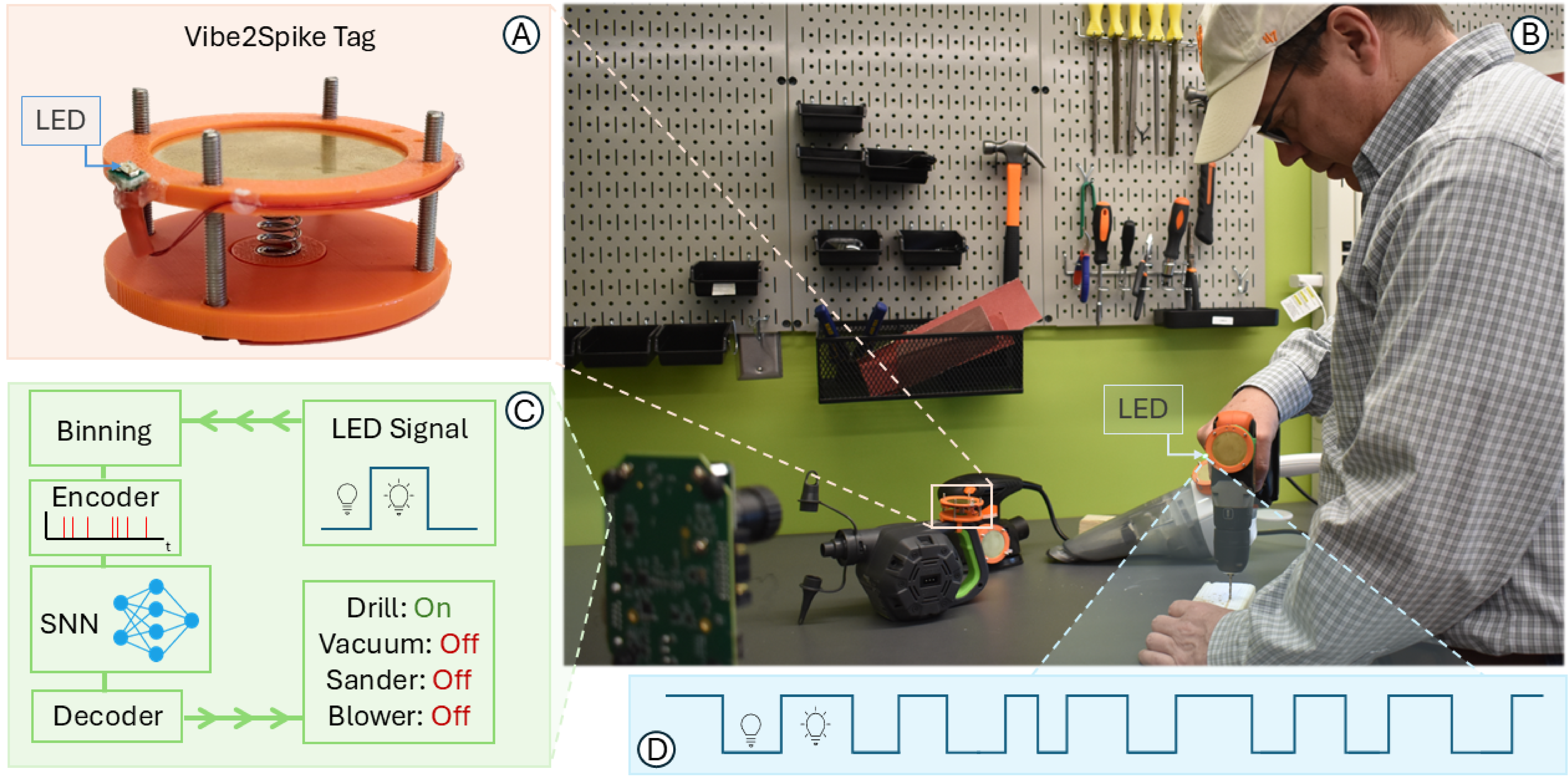} 
  \caption{Vibe2Spike sensor deployed in workshop on multiple devices. B) Vibe2Spike sensor attached to a drill emitting LED flashes based on the vibrations of the device. C) LED on/off signal encoded as spikes sent to the SNN network are decoded and classified by device type. D) The temporal LED events are created by the on/off flashing of the LED producing unique per device signals.}
  \label{fig:teaser}
\end{figure*}

\IEEEpubidadjcol
\begin{abstract}
The deployment of dense, low-cost sensors is critical for realizing ubiquitous smart environments. However, existing sensing solutions struggle with the energy, scalability, and reliability trade-offs imposed by battery maintenance, wireless transmission overhead, and data processing complexity. In this work, we present \textit{Vibe2Spike}, a novel battery-free, wireless sensing framework that enables vibration-based activity recognition using visible light communication (VLC) and spiking neural networks (SNNs). Our system uses ultra-low-cost tags composed only of a piezoelectric disc, a Zener diode, and an LED, which harvest vibration energy and emit sparse visible light spikes without requiring batteries or RF radios. These optical spikes are captured by event cameras and classified using optimized SNN models evolved via the EONS framework. We evaluate \textit{Vibe2Spike} across five device classes, achieving 94.9\% average classification fitness while analyzing the latency-accuracy trade-offs of different temporal binning strategies. \textit{\textit{Vibe2Spike}} demonstrates a scalable, and energy-efficient approach for enabling intelligent environments in a batteryless manner.
\end{abstract}


\section{Introduction}
The promise of ubiquitous smart environments—industrial workshops predicting machinery faults, kitchens monitoring appliance usage, or buildings diagnosing structural wear—remains hindered by the energy-cost-reliability trade-offs of conventional sensing systems. While vibrations offer rich insights into machine health, human activity, etc., deploying sensors at scale requires solutions that are batteryless (to eliminate maintenance), wireless (to simplify installation), and ultra-low-cost (to enable dense deployments). Existing systems fall short in both sensing and computation. Battery-powered IoT sensors waste energy on continuous RF transmissions and require regular maintenance, while batteryless RF backscatter tags suffer from limited range (1–5m) and reduced reliability in RF-congested environments. On the computational side, conventional CNN-based edge inference is energy-intensive and poorly matched to sparse, event-driven data streams. Although spiking neural networks (SNNs) offer promise for low-power, temporal processing, they remain largely confined to simulation and have yet to be widely deployed with real-world sensing hardware.

Overcoming these limitations requires rethinking the sensing, communication, and computation layers together. First, sensing must be battery-free and direct. We propose a vibration-to-spike encoding approach, where ambient mechanical vibrations are harvested and directly converted into sparse optical pulses—eliminating the need for bulky batteries, microcontrollers, or complex electronics. Second, wireless communication must be long-range, scalable, and resilient to interference. Traditional RF systems suffer from congestion \cite{globe_com_congestion, mobihoc}, limited range under low-power operation, and spectrum saturation from widely deployed wireless technologies such as Wi-Fi, Bluetooth Low Energy (BLE), etc. In contrast, visible Light Communication (VLC) offers a compelling alternative, as it is completely immune to interference from the RF spectrum, making it especially well-suited for dense sensing environments. Importantly for this work, VLC pairs naturally with neuromorphic (event-based) cameras, which are optimized to detect sparse optical events with high temporal precision and wide dynamic range. This makes them ideally suited to capture the brief, low-power light pulses emitted by our batteryless vibration tags, even at long distances \cite{dan_paper}. Finally, the computation layer must align with the sparse, event-driven nature of the signal. Frame-based or dense neural networks are inefficient in this regime. In contrast, Spiking Neural Networks (SNNs) operate naturally on asynchronous inputs and support real-time, ultra-low-power inference. In our system, SNNs are deployed at the receiver—co-located with the neuromorphic camera—to classify incoming optical spike streams and infer machine or activity states. Despite advances in each of these domains individually, no prior work integrates batteryless sensing, VLC-based communication, and SNN-driven neuromorphic computation into a unified system that addresses the combined challenges of power, scalability, and security in vibration sensing.

To bridge this gap, we propose \textit{\textit{Vibe2Spike}}: a novel battery-free wireless sensing framework for vibration activity detection using spiking neural networks. Our solution leverages low-cost sensor tags (less than \$1) composed of a piezoelectric disc (piezo), a diode, and an LED. These tags harvest vibrational energy and emit visible light pulses ("spikes") without needing external power sources or traditional wireless radios. Low-power event cameras capture these optical spikes, enabling sparse, efficient sensing and communication.

Building on this, we employ the evolutionary optimization for neuromorphic systems framework (EONS) \cite{schuman2020evolutionary} to design SNNs that classify vibration-induced optical signals, allowing machine or activity recognition. Unlike prior VLC systems focused on high-bandwidth data, we optimize for ultra-low-power, event-driven transmission that matches the sparse output of both the sensor and the neuromorphic processor.

By tightly integrating batteryless vibration-to-spike encoding, visible light communication, and neuromorphic edge processing, \textit{Vibe2Spike} achieves, to the best of our knowledge, the first practical realization of SNNs for battery-free vibration sensing. This integration substantially lowers power requirements, simplifies hardware, enhances scalability, enabling smart environments capable of predictive maintenance, automated machine recognition, and context-aware interactions at minimal cost.

Our Contributions include:

\begin{itemize}
    \item \textbf{Battery-Free, \textit{Vibe2Spike} Tag:} We develop a low-cost, self-powered tag that converts ambient vibrations into sparse visible light pulses using only a piezo, diode, and LED, without batteries, RF modules, or microcontrollers.
    
    \item \textbf{End-to-End Neuromorphic Sensing Pipeline:} We design an end-to-end event-based pipeline, directly coupling physical spike generation with temporally binned SNN inference, eliminating frame-based processing.
    
    \item \textbf{Evolutionary SNN Optimization:} We apply the EONS framework to evolve lightweight SNN architectures optimized for accuracy, latency, and hardware efficiency, enabling classification of five vibration sources.
    
    \item \textbf{Empirical Evaluation:} We achieve 94.9\% average classification fitness across five classes in a workshop environment where we track equipment activity and empirically evaluate system performance in situ.
\end{itemize}
\section{Related Work}

\textbf{Battery-Free Sensing:} Battery-free sensing systems have evolved from early RFID-based tags like WISP~\cite{sample2008design} and Moo~\cite{moo}, which used RF backscatter to transmit sensor data with harvested energy. Hybrid analog-digital schemes (e.g., Talla et al.\cite{talla2013hybrid}) enabled acoustic sensing via WISP, while later advancements like PaperID\cite{katsuragawa2019tip}, RF Bandaid~\cite{ranganathan_rf_2018}, and LiveTag~\cite{gao2019livetag} expanded to touch and environmental interactions. Despite progress, RF-based systems face inherent trade-offs: passive tags achieve less than 12m range (e.g., RF Bandaid’s 9m at 160µW~\cite{ranganathan_rf_2018}), while active LoRa backscatter~\cite{Talla2017Sep} extends range to hundreds of meters but requires dedicated infrastructure and risks eavesdropping due to omnidirectional signal leakage. Non-backscatter approaches like Pible~\cite{fraternali_pible_2018} (648µW BLE) and luXbeacon~\cite{jeon_luxbeaconbatteryless_2019} (100m range) reduce deployment complexity but demand energy-intensive radios, limiting scalability.

\textbf{Visible Light Communication (VLC):} VLC systems have primarily targeted low-power data transmission but face constraints due to camera and modulation inefficiencies. Early systems like LED-to-LED communication~\cite{schmid_led_2013} achieved modest ranges of up to 2m at 289mW. Successive innovations, including RetroVLC~\cite{li_retro-vlc_2015} and PassiveVLC~\cite{xu_passivevlc_2017}, significantly reduced power consumption to 90µW and 150µW, respectively but remained limited to short-range communication (2.4m and 4.5m). Subsequent solutions, such as RetroTurbo~\cite{wu_turboboosting_2020} achieved an improved range of 7.5m (0.8mW  8kbps), and RetroMUMIMO~\cite{xu_low-latency_2022} enabled concurrent transmissions at up to 3.75m. 

Although LightAnchors~\cite{lightanchors2019uist} effectively utilized smartphone cameras operating at 240 fps for LED detection, its reliance on traditional RGB cameras ($\geq$ 500 mW) resulted in high power demands for continuous operation. Additionally, prior VLC systems target high-rate communication rather than sparse, low-power event transmission. Our work fills this gap by using VLC for sparse, event-driven communication of vibration events, optimized for low energy budget and real-time decoding with neuromorphic processors.

\textbf{VLC as an Alternative to RF-based Sensing:} Our work, \textit{Vibe2Spike}, builds on these foundations and addresses the limitations of both RF and VLC-based systems by tightly integrating VLC with neuromorphic sensing and edge processing. First, VLC enables us to avoid RF spectrum congestion—a growing concern due to widespread use of WiFi, BLE, and other wireless technologies. Unlike RF backscatter systems that expose transmissions broadly and risk eavesdropping, our system confines communication within a physical space using visible light. VLC signals cannot penetrate walls, and with highly directional LEDs, we constrain signal propagation to the reader's field of view, enhancing privacy and security. In contrast to RF-based systems that require complex security protocols~\cite{barman_every_2021}, users of \textit{Vibe2Spike} can physically disable communication (e.g., covering the LED) for intuitive, low-cost privacy control~\cite{do2023powering}. Finally, VLC introduces no electromagnetic interference, making \textit{Vibe2Spike} suitable for sensitive environments like hospitals and EMI-heavy industrial settings. By combining these communication advantages with battery-free sensing and SNN-based neuromorphic processing, our system sets a new direction for low-power, secure, and scalable smart environment sensing.

\textbf{Edge AI with Spiking Neural Networks (SNNs):} In addition to innovations in sensing and communication, \textit{Vibe2Spike} advances the computational layer by leveraging energy-efficient neuromorphic inference through Spiking Neural Networks (SNNs). SNNs offer a promising path for energy-efficient edge AI due to their sparse, event-driven nature. Applications include sub-µW biosignal decoding~\cite{de2025neuromorphic, rivelli2025adaptively}, event-based vision~\cite{wu2024lightweight}, and distributed low-bandwidth sensor networks~\cite{chen2024neuromorphic}. However, most existing work assumes stable power availability, often simulating SNNs on neuromorphic hardware without integrating them with energy-constrained sensors. Moreover, evolutionary frameworks like EONS~\cite{schuman2020evolutionary} optimize SNN architectures for static datasets but have not fully explored end-to-end deployment with real-world, batteryless, event-driven sensors. Our work bridges this gap by coupling physical spike generation from a vibration-powered tag with evolutionary-optimized SNNs for activity classification. This constitutes, to the best of our knowledge, the first integration of battery-free sensing, VLC, and neuromorphic processing into a unified, deployable system.

\section{System Overview}
We now describe the components of the \textit{Vibe2Spike} system, illustrated in Fig.~\ref{fig:teaser}. \textit{Vibe2Spike} is designed to detect, process, and classify the unique vibration patterns of human-operated devices in real time, enabling the transformation of everyday environments—such as workshops and kitchens—into smart, context-aware spaces. These settings are rich in vibration signals generated by tools and appliances, but current sensing technologies either ignore these signals or are too costly and power-hungry for scalable deployment. The first component is the \textit{Vibe2Spike} tag (Fig.~\ref{fig:teaser}A), a low-cost, battery-free sensor that converts mechanical vibrations into visible light pulses. Costing less than \$1, each tag is composed of a piezoelectric disc, a diode, and an LED. It harvests vibrational energy to power the LED, producing sparse optical spikes. The tag can be attached to any device that vibrates—such as a sander, blower, or vacuum—providing a lightweight and unobtrusive sensing mechanism. Each device’s vibration generates a characteristic temporal signature in the emitted light. A neuromorphic event-based camera (Fig.~\ref{fig:teaser}B) captures these light pulses as asynchronous ON/OFF events, enabling high-speed, low-power sensing. To prepare the event stream for classification, we implement a temporal binning pipeline that groups events into fixed-size time windows and converts them into structured feature vectors. This process preserves the temporal structure of the signal while making it compatible with learning algorithms. Finally, a Spiking Neural Network (SNN) processes these feature vectors using an encoder-decoder architecture. It classifies the input based on the temporal dynamics of each device’s vibration signature, enabling real-time, energy-efficient recognition of equipment activity.

\begin{figure}
    \centering
    \includegraphics[width=1\linewidth]{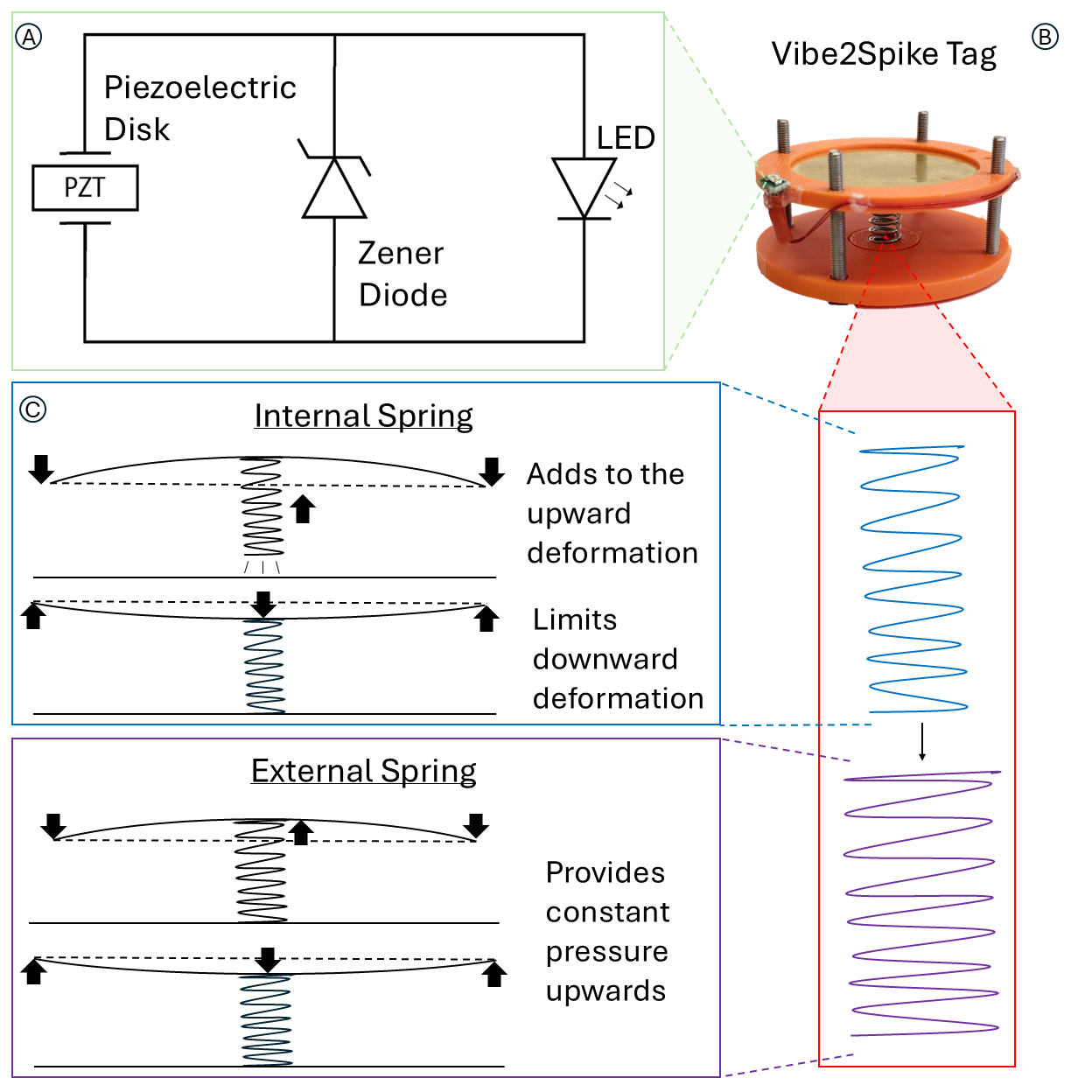}
    \caption{
        \textbf{\textit{Vibe2Spike} schematic.} 
        (A) Electrical circuit diagram showing a piezoelectric disk, a Zener diode, and an LED connected in parallel to convert vibration into visible light pulses. 
        (B) Assembled \textit{Vibe2Spike} tag with 3D-printed housing and embedded dual-spring mechanism. 
        (C) Cross-sectional mechanical diagram illustrating the external spring (surrounding) for preload tension and the internal spring (centered) for asymmetric deformation under vibration.
    }
    \label{fig:schematic}
\end{figure}


\subsection{\textit{Vibe2Spike} Tag Design}
We describe the design and function of the \textit{Vibe2Spike} tag, illustrated in Fig.~\ref{fig:schematic}. The tag's primary function is to convert ambient mechanical vibrations into visible light pulses in a battery-free, power-efficient manner—enabling scalable, low-cost sensing in environments such as workshops and kitchens, where vibration signals are abundant. Each tag consists of a 40mm piezoelectric disk (Same Sky CEB-44D06 or equivalent), a surface-mounted LED (Kingbright APT3216LSECK/J4-PRV), a 0.5W 12V Zener diode, and a 3D-printed case with a dual-spring mechanism. The vibration-to-light conversion circuit is shown in Fig.~\ref{fig:schematic}A. Two wires—one from the center electrode and one from the outer metal shim—connect the piezo to the LED and Zener diode in parallel. The Zener diode regulates output voltage, protects the LED from surges, and rectifies the AC signal, ensuring current flows only when forward-biased.

Because the raw current generated by typical piezo deformation is insufficient to power an LED directly, the tag relies on \textit{mechanical amplification} rather than electronic amplification. The assembled tag is shown in Fig.~\ref{fig:schematic}B, while the dual-spring design is detailed in Fig.~\ref{fig:schematic}C. The \textbf{external spring} (bottom of Fig.~\ref{fig:schematic}C) is positioned beneath the piezo and surrounds the inner spring concentrically. It applies a constant upward preload, keeping the piezo pressed against the housing and ensuring mechanical tension is maintained during operation. Nested inside it is the \textbf{internal spring} (top of Fig.~\ref{fig:schematic}C), located directly below the piezo's center. This inner spring is stiffer and more constrained in movement. It is designed to permit greater upward flexing while limiting downward displacement, thus introducing asymmetry in the piezo’s strain response.

As the tag experiences oscillatory motion from the device it's mounted on, the piezo’s perimeter—rigidly clamped to the case—follows the device’s vibration, while the center flexes differentially due to the internal spring. This causes a large deformation gradient across the piezo disk. The resulting alternating strain produces an AC voltage signal. When this voltage exceeds the LED’s forward threshold, the circuit emits a visible light pulse. The Zener diode ensures clean, unidirectional current flow, and the blinking pattern mirrors the source vibration frequency. To assemble the tag, the internal spring is placed in the central spring cavity of the lower case, directly under the piezo center. The external spring is placed concentrically around it. The piezo is clamped between the top and middle layers using two short screws, then mounted to the bottom piece with four longer screws. The springs are compressed to achieve a total tag height of 13.8–12.8,mm, ensuring consistent mechanical preload.

In summary, the \textit{Vibe2Spike} tag uses a nested dual-spring design and a passive circuit to convert vibration into optical pulses—without batteries, microcontrollers, or wireless radios. Its simplicity and low power requirements make it suitable for dense, cost-effective deployment in real-world smart environments.

\subsection{Event-Based Vibration Capture}

As the next step in our system, the blinking LED output from the deployed \textit{Vibe2Spike} tag is captured by a neuromorphic, event-based camera, which records changes in brightness as a sequence of discrete ON (1) and OFF (0) events. Unlike conventional frame-based cameras, event-based sensors detect only pixel-wise changes in intensity, enabling microsecond-level temporal resolution and sub-millisecond latency~\cite{chakravarthi_recent_2024}. For our system, we use the EVK3HD camera from Prophesee, which supports asynchronous event detection with minimal power consumption.

As the LED flickers in response to device vibrations, multiple adjacent pixels in the camera's field of view are triggered. Each brightness change generates an event tuple:

\textbf{Event = (x, y, polarity, timestamp)}

\begin{itemize}
\item coordinate \textit{(x, y)} denotes the pixel location,
\item \textit{polarity} is 1 for ON events and 0 for OFF events,
\item \textit{timestamp} provides microsecond-resolution timing.
\end{itemize}

These asynchronous events collectively encode the temporal signature of a device's vibrations.

\subsection{Event Data Preprocessing}

To ensure high-quality input during training, we performed a 1-minute calibration phase to identify the pixel with the highest cumulative event count as the source of the temporal signal for each device. This design choice reduced noise and improved label fidelity, extracting high-SNR temporal features to train the Spiking Neural Network (SNN), enabling us to focus on the core challenge of designing end-to-end event-driven neuromorphic sensing using SNNs. While this heuristic aids training, our trained model is agnostic to pixel location and can classify vibration events from any pixel's temporal stream, the overall pipeline is compatible with broader pixel-parallel architectures. In prior work (NeuroCamTag~\cite{scott_neurocamtags_2024}), we developed streaming event queuing and pixel-wise processing methods suitable for scaling to full-field inference. These techniques can be integrated into future versions of \textit{Vibe2Spike} to support real-time classification across all active pixels without requiring a fixed localization step.

\subsection{Temporal Event Binning and Feature Extraction}\label{binning}

To transform raw event streams into structured input for classification, we apply a temporal binning process. Events from the selected LED pixel are grouped into fixed-size time windows of 50 milliseconds. Within each bin, we compute the total number of ON events (brightness increase) and OFF events (brightness decrease), resulting in a two-dimensional feature per bin.

For a sequence of \textit{N} consecutive bins, we concatenate the ON/OFF counts to form a 2$\times$\textit{N} feature vector. This representation preserves the temporal structure of the signal while reducing data complexity for downstream learning. For instance:
\begin{itemize}
    \item \textit{N} = 10 bins → 0.5s total duration → 20 input features
    \item \textit{N} = 50 bins → 2.5s total duration → 100 input features
\end{itemize}

In our evaluation, by varying \textit{N}, we investigate the trade-off between classification latency and accuracy: shorter sequences offer faster inference, while longer ones may capture more stable temporal patterns. These feature vectors are then used to train the SNN, as described in the next section.

\subsection{Spiking Neural Network Design and Training}\label{snndesign}
Spiking Neural Networks (SNNs) are especially well-suited for the \textit{Vibe2Spike} system because they align with the key properties of our sensing pipeline: sparse, asynchronous, and temporally structured data. Unlike traditional neural networks that rely on dense, frame-based inputs, SNNs operate on discrete spikes, making them a natural fit for processing the binary ON/OFF events generated by our event-based camera in response to vibration-induced LED flickers.

This spike-based processing allows SNNs to efficiently capture the temporal dynamics of each device’s vibration signature—critical for distinguishing between tools that exhibit overlapping frequency bands or intermittent operation. Furthermore, because SNNs require significantly fewer operations than conventional CNNs and can be deployed on neuromorphic hardware, they are ideal for real-time inference in energy-constrained environments like workshops and kitchens, where \textit{Vibe2Spike} tags are deployed.

To classify device activity based on temporally binned event features, we employ a spiking neural network (SNN) trained using the TENNLab neuromorphic framework~\cite{psb:18:ten} and used the the Evolutionary Optimization of Neuromorphic Systems (EONS)~\cite{schuman_evolutionary_2020} to optimize the SNNs.  EONS uses a genetic algorithm to evolve populations of SNNs over multiple generations. Each network is evaluated for classification accuracy and undergoes mutation, crossover, and selection to form the next generation. This enables joint optimization of both the network’s topology (e.g., number of neurons, recurrent loops, sparsity) and parameters (e.g., synaptic weights, time constants). Besides the network itself, when developing spiking neural networks, input data must be encoded into spikes and output spikes from the network must be decoded into some decision or classification.


\subsubsection{Input Encoding: Argyle-4 Scheme}

Following the binning of event data (as described in Section~\ref{binning}), we extract temporal features consisting of ON and OFF event counts. These features are then encoded into spikes using the Argyle-4 encoding scheme~\cite{spb:19:nte,prw:25:cpa}. Our choice of encoding technique is based on both manual hyperparameter experimentation and prior applications in neuromorphic control and regression~\cite{prw:25:cpa,pdg:24:risp}.

\begin{figure}[htbp]
\centering
\includegraphics[width=3in]{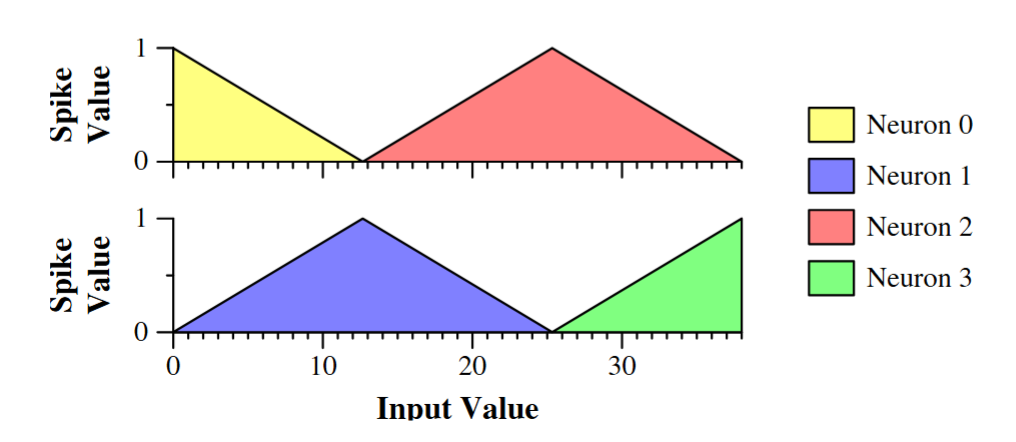}
\caption{\label{fig:argyle} Diagram to show how an argyle encoder converts values from 0 to
38 into two spikes having values between 0 and 1.}
\end{figure}

For each experimental run, the bin size determines the resolution of temporal features and hence the spike encoding granularity. For example, if we use a bin size of 50, we obtain 50 ON-event counts and 50 OFF-event counts—yielding 100 input values per sample. These values are automatically scaled into a normalized range, based on the minimum and maximum values observed in the training data. While the example shown in Fig.~\ref{fig:argyle} uses a 0–38 range, this range is dataset-dependent and determined dynamically.

Each SNN input sample contains 100 temporal features (for binsize=50) derived from 50 ON and 50 OFF event counts. These values are then encoded into spikes using the Argyle-4 scheme. The encoding proceeds by processing each ON/OFF pair every three simulation timesteps. Each input value produces two spikes with complementary magnitudes $x$ and $1 - x$, and thus each pair results in four spikes total. These spikes are applied to a subset of input neurons in each 3-timestep window. With 50 such pairs, the encoding spans 150 timesteps and injects exactly 200 spikes into the network per sample.

The Argyle-4 encoder partitions the input space into equal-sized regions and maps each value to a complementary pair of spikes applied to adjacent input neurons. This “charge injection” technique conserves input energy, allows smooth transitions between neighboring values, and is efficient for neuromorphic deployment.

This encoding strategy (for our dataset), visualized in Fig.\ref{fig:argyle}, provides a structured and reliable way to inject temporally binned event features into a spiking neural network. It has demonstrated effectiveness across multiple neuromorphic tasks, including CartPole control\cite{prw:25:cpa} and function approximation~\cite{pdg:24:risp}, and aligns well with the sparse, time-varying signals produced by our event-based sensing pipeline.

\subsubsection{Output Decoding}

After spike encoding, each input sample is processed by a spiking neural network (SNN) trained to classify the device based on its vibration signature. We employ the \textbf{Winner-Take-All (WTA)} decoding strategy, commonly used in neuromorphic classification. Each output neuron corresponds to one of the device classes. During inference, the output neuron that emits the highest number of spikes over the simulation window determines the predicted class. This approach is simple, interpretable, and compatible with the temporal dynamics of spiking activity.

\section{Evaluation and Results}
To evaluate the effectiveness of the \textit{Vibe2Spike} system, we conducted experiments using a representative set of human-operated vibratory tools. Our goal was to assess the system's ability to distinguish between devices based on their vibration-induced light signals, captured via an event camera and classified by a spiking neural network.

\subsection{Device Selection and Experimental Setup}

\begin{figure}[ht]
\centering
\includegraphics[width=1\linewidth]{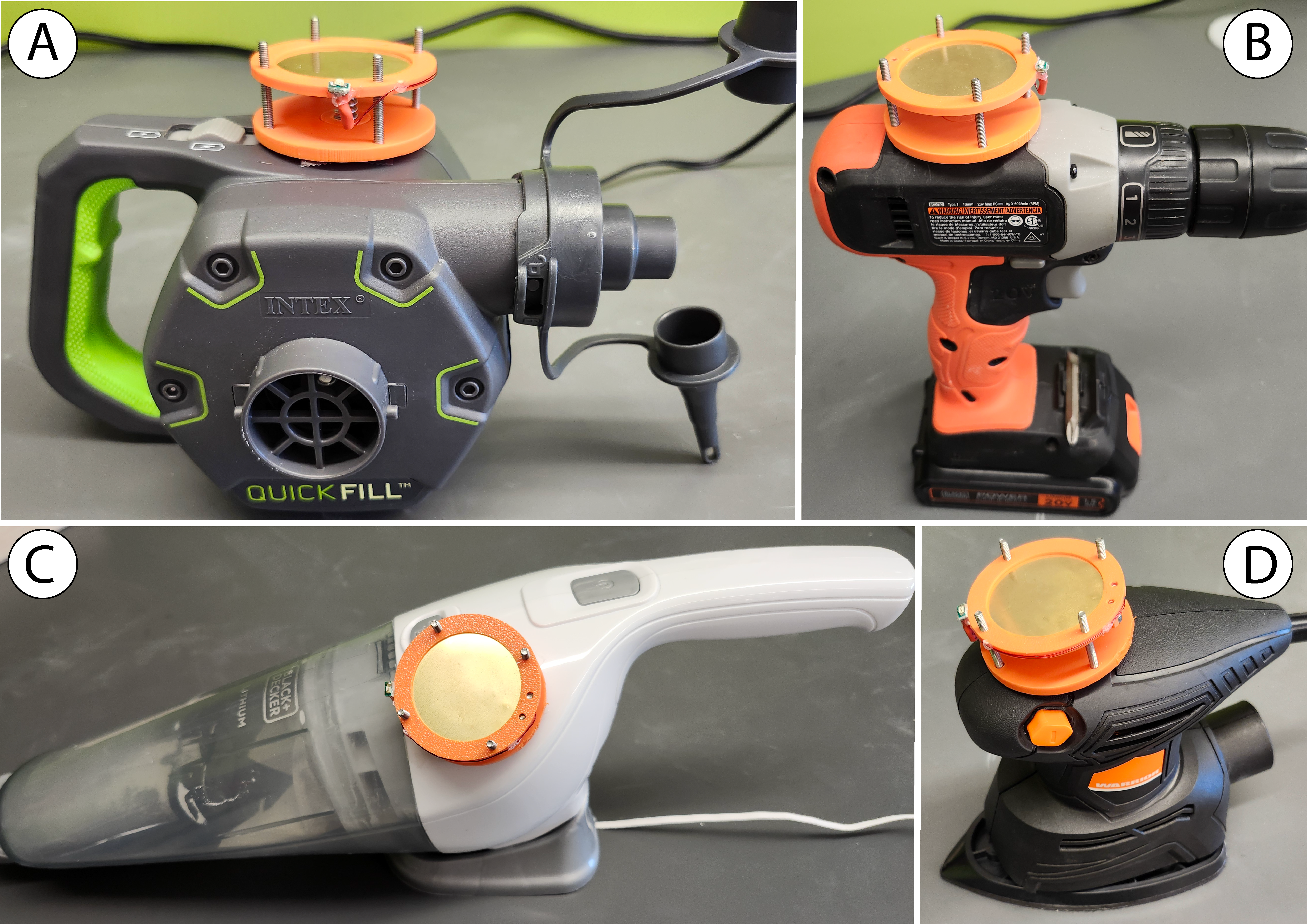}
\caption{Four common handheld devices used in evaluation: (A) inflator blower, (B) cordless drill, (C) handheld vacuum, and (D) palm sander.}
\label{fig:alldevices}
\end{figure}

We selected four common devices with differing mechanical vibration signatures (Fig.~\ref{fig:alldevices}): a palm sander, a cordless drill, a handheld vacuum, and an inflator blower. These tools vary in frequency and duty cycle characteristics and include devices that exhibit little to no visible movement, making them ideal candidates to evaluate the effectiveness of a vibration-only sensing method.

Each device was fitted with a \textit{Vibe2Spike} tag and operated under standard conditions. The event-based camera was placed at a fixed location approximately 30 cm away from the device, with a clear line-of-sight to the \textit{Vibe2Spike} tag’s LED. Based on our prior work with NeuroCamTag~\cite{scott_neurocamtags_2024}, we found that the LED remains reliably detectable by the event camera even at oblique angles, with up to 85 degrees of angular offset between the LED and the camera’s optical axis. This tolerance ensured consistent event capture across all trials, even when slight variations occurred in device orientation. We collected five trials per device, yielding a total of 25 minutes of labeled event data across all classes.

\subsection{Event Volume and Feature Distributions}

\begin{figure}
\centering
\includegraphics[width=1\linewidth]{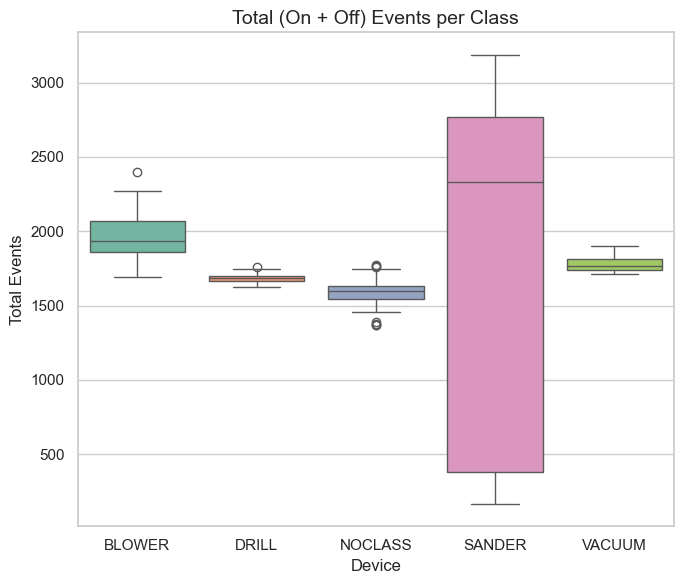}
\caption{Event count distributions across different vibration classes.}
\label{fig:boxplot}
\end{figure}

Due to differing vibration intensities, device mass, and mounting configurations, the number of ON/OFF events generated by the \textit{Vibe2Spike} tag varies significantly across devices. Fig.~\ref{fig:boxplot} shows the distribution of total events per device class. For instance, the sander consistently produces more events, while the vacuum and inflator yield fewer due to less aggressive vibrations.

While these differences influence the raw signal volume, the SNN must learn to distinguish devices based not only on count but on the \textit{temporal structure} of the events. This figure demonstrates that classification is not trivially tied to event count—devices like the drill and blower exhibit overlapping distributions—indicating that the SNN is learning nuanced, time-dependent patterns rather than simply relying on total event magnitude.




\subsection{Classification Experiment Design}

To evaluate classification performance, we used the full \textit{Vibe2Spike} pipeline—including temporal binning (Section~\ref{binning}), Argyle-4 spike encoding and SNN classification (Section~\ref{snndesign})—in conjunction with the EONS evolutionary training framework. For each binning configuration, we generated a population of 999 randomly initialized spiking neural networks and trained them over multiple generations using classification accuracy as the fitness metric. We tested four bin sizes, corresponding to latency windows ranging from 250 ms to 5 seconds. All models were validated on holdout samples, and the best-performing model for each configuration is reported in Table~\ref{validation}.

\subsection{Evaluation Metrics and Confusion Matrix}

Table~\ref{validation} shows that a 2,500 ms (2.5-second) binning window yields the highest F1 accuracy of 94.9\%, outperforming shorter or longer durations. This suggests that coarser temporal bins effectively aggregate sparse optical spikes while reducing sensitivity to momentary fluctuations. Shorter bins may fragment the signal, whereas longer bins preserve the characteristic signal profile.

\begin{figure}
    \centering
    \includegraphics[width=1\linewidth]{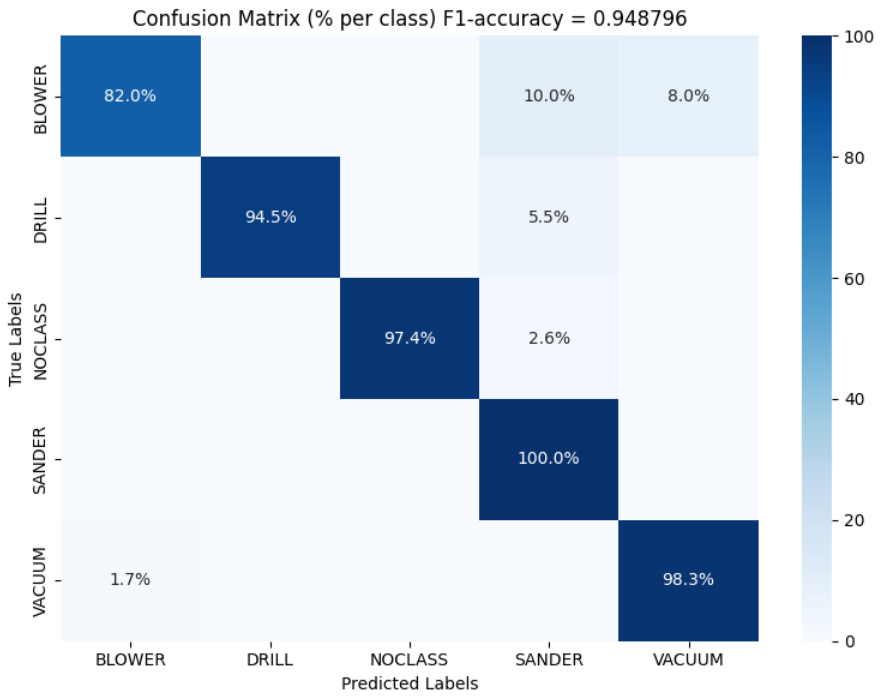}
    \caption{Confusion matrix of the five classes from the best SNN network with a F1 accuracy of 94.9\%.}
    \label{fig:confusionmatrix}
\end{figure}

The confusion matrix in Fig.\ref{fig:confusionmatrix} confirms strong per-class accuracy, with only minor confusion between blower and vacuum devices—two classes with similar frequency profiles. These results highlight the critical role of binning in balancing latency, noise robustness, and signal expressiveness in neuromorphic pipelines like \textit{Vibe2Spike}.

\subsection{Pipeline Latency and Design Trade-offs}

The temporal binning strategy introduces an important trade-off between latency and classification accuracy. Table~\ref{validation} summarizes this relationship across four binning configurations. With only five bins (250 ms total latency), the system responds quickly but achieves lower F1 accuracy (81.16\%). As the bin count increases, classification accuracy improves, peaking at 94.88\% with 50 bins (2.5s latency). However, further increasing to 100 bins (5s) yields only marginal gains while introducing additional delay.

These results demonstrate the flexibility of the \textit{Vibe2Spike} pipeline: developers can tune bin size to prioritize speed (e.g., for real-time alerts) or precision (e.g., for post-hoc analysis). While longer temporal windows allow better aggregation of signal structure, they also delay response time, which may not be suitable for time-critical applications.

\begin{table}[h]
\centering
\caption{F1-Accuracy Validation Scores}
\label{validation}
\begin{tabular}{|c|c|c|}
\hline
Latency Time & Bins & Validation (\%)  \\ \hline
250 ms & 5    & 81.16                  \\ \hline
500 ms & 10   & 88.92                  \\ \hline
2,500 ms & 50   & \textbf{94.88}       \\ \hline
5,000 ms & 100  & 93.52                \\ \hline
\end{tabular}
\end{table}

\section{Discussion}
The results demonstrate that \textit{Vibe2Spike} achieves strong classification performance using a minimalist, battery-free sensing architecture. With an average F1 accuracy of 94.9

The effectiveness of longer temporal bins (e.g., 2.5s) observed in our experiments points to a broader design consideration: binning is not merely a preprocessing step, but a critical tuning knob that governs the balance between responsiveness and reliability in event-driven pipelines. For applications where immediate decisions are not essential, coarser bins can enhance signal quality and stability. This reinforces the suitability of spiking neural networks, which are naturally aligned with sparse, temporally structured input and avoid the inefficiencies of dense, frame-based deep learning models.

Unlike traditional wireless sensing systems that rely on RF radios, microcontrollers, and batteries, \textit{Vibe2Spike} eliminates the need for active power sources or wireless chips. The tag leverages visible light communication (VLC) and harvests energy from ambient vibrations, reducing cost, maintenance burden, and power consumption. Compared to prior VLC-based systems, our approach offers significantly lower hardware complexity and leverages the neuromorphic properties of event cameras to produce energy-efficient, high-temporal-resolution input streams.

Nevertheless, limitations remain. The system’s performance depends on the strength and consistency of the device's vibrations, which may vary across use contexts. Temporal binning introduces latency that may be unsuitable for applications requiring sub-second response times. Moreover, while the Evolutionary Optimization of Neuromorphic Systems (EONS) framework yields highly efficient SNN models, its training process is computationally intensive and sensitive to parameter tuning, which could limit rapid deployment in new settings.

Another key design consideration in this work was our decision to isolate the LED signal by selecting the pixel with the highest cumulative event count during training. This heuristic was used to improve label fidelity and provide the cleanest signal for learning, but it is not a requirement at inference time. Our trained SNN generalizes across pixels and is compatible with a broader, pixel-parallel pipeline. In future iterations of \textit{Vibe2Spike}, we plan to support full-field, real-time classification by incorporating high-throughput event queuing and pixel-wise processing techniques, such as those developed in our NeuroCamTag system~\cite{scott_neurocamtags_2024}.

Finally, we envision future improvements across three axes: (1) extending \textit{Vibe2Spike} to function robustly under varying lighting and occlusion conditions, (2) enabling multi-tag scenarios with overlapping optical signals and real-time source separation, and (3) deploying on neuromorphic hardware accelerators for edge inference. These directions will help realize a scalable, ultra-low-power, event-driven sensing system for context-aware environments.

\section{Conclusion}

In conclusion, \textit{Vibe2Spike} presents a novel, battery-free, and wireless sensing framework that combines visible light communication with spiking neural networks to enable vibration-based activity recognition in smart environments. By using ultra-low-cost, passive tags composed solely of a piezoelectric disc, a Zener diode, and an LED, the system achieves scalable and reliable sensing without batteries or RF radios. Our evaluation across five device classes demonstrates a high classification accuracy of 94.9\%, underscoring the effectiveness of our event-driven pipeline. Additionally, our analysis of latency-accuracy trade-offs offers important design insights for real-time neuromorphic sensing applications. \textit{Vibe2Spike} represents a promising advance toward sustainable, intelligent sensing infrastructure. Future work will focus on scaling to larger deployments, integrating additional energy-harvesting modalities, and improving SNN training and hardware deployment strategies.optimization techniques.

\bibliographystyle{IEEEtran}
\bibliography{refs.bib}

\begin{thebibliography}{10}
\providecommand{\url}[1]{#1}
\csname url@samestyle\endcsname
\providecommand{\newblock}{\relax}
\providecommand{\bibinfo}[2]{#2}
\providecommand{\BIBentrySTDinterwordspacing}{\spaceskip=0pt\relax}
\providecommand{\BIBentryALTinterwordstretchfactor}{4}
\providecommand{\BIBentryALTinterwordspacing}{\spaceskip=\fontdimen2\font plus
\BIBentryALTinterwordstretchfactor\fontdimen3\font minus \fontdimen4\font\relax}
\providecommand{\BIBforeignlanguage}[2]{{%
\expandafter\ifx\csname l@#1\endcsname\relax
\typeout{** WARNING: IEEEtran.bst: No hyphenation pattern has been}%
\typeout{** loaded for the language `#1'. Using the pattern for}%
\typeout{** the default language instead.}%
\else
\language=\csname l@#1\endcsname
\fi
#2}}
\providecommand{\BIBdecl}{\relax}
\BIBdecl

\bibitem{globe_com_congestion}
\BIBentryALTinterwordspacing
Q.~D. La, D.~Nguyen-Nam, M.~V. Ngo, and T.~Q.~S. Quek, ``Coexistence evaluation of densely deployed ble-based body area networks,'' in \emph{GLOBECOM 2017 - 2017 IEEE Global Communications Conference}.\hskip 1em plus 0.5em minus 0.4em\relax IEEE Press, 2017, p. 1–6. [Online]. Available: \url{https://doi.org/10.1109/GLOCOM.2017.8253947}
\BIBentrySTDinterwordspacing

\bibitem{mobihoc}
\BIBentryALTinterwordspacing
H.~Qiu, K.~Psounis, G.~Caire, K.~M. Chugg, and K.~Wang, ``High-rate wifi broadcasting in crowded scenarios via lightweight coordination of multiple access points,'' in \emph{Proceedings of the 17th ACM International Symposium on Mobile Ad Hoc Networking and Computing}, ser. MobiHoc '16.\hskip 1em plus 0.5em minus 0.4em\relax New York, NY, USA: Association for Computing Machinery, 2016, p. 301–310. [Online]. Available: \url{https://doi.org/10.1145/2942358.2942372}
\BIBentrySTDinterwordspacing

\bibitem{dan_paper}
\BIBentryALTinterwordspacing
D.~Scott, M.~Bringle, I.~Fahad, G.~Morales, A.~Zahid, and S.~Swaminathan, ``Neurocamtags: Long-range, battery-free, wireless sensing with neuromorphic cameras,'' \emph{Proc. ACM Interact. Mob. Wearable Ubiquitous Technol.}, vol.~8, no.~3, Sep. 2024. [Online]. Available: \url{https://doi.org/10.1145/3678529}
\BIBentrySTDinterwordspacing

\bibitem{schuman2020evolutionary}
C.~D. Schuman, J.~P. Mitchell, R.~M. Patton, T.~E. Potok, and J.~S. Plank, ``Evolutionary optimization for neuromorphic systems,'' in \emph{Proceedings of the 2020 Annual Neuro-Inspired Computational Elements Workshop}, 2020, pp. 1--9.

\bibitem{sample2008design}
A.~P. Sample, D.~J. Yeager, P.~S. Powledge, A.~V. Mamishev, and J.~R. Smith, ``Design of an rfid-based battery-free programmable sensing platform,'' \emph{IEEE transactions on instrumentation and measurement}, vol.~57, no.~11, pp. 2608--2615, 2008.

\bibitem{moo}
\BIBentryALTinterwordspacing
U.~of~Massachusetts~Amherst. (2023) Unichmoo. University of Massachusetts Amherst. [Online]. Available: \url{https://github.com/spqr/umichmoo}
\BIBentrySTDinterwordspacing

\bibitem{talla2013hybrid}
V.~Talla, M.~Buettner, D.~Wetherall, and J.~R. Smith, ``Hybrid analog-digital backscatter platform for high data rate, battery-free sensing,'' in \emph{2013 IEEE Topical Conference on Wireless Sensors and Sensor Networks (WiSNet)}, IEEE.\hskip 1em plus 0.5em minus 0.4em\relax Unknown: IEEE, 2013, pp. 1--3.

\bibitem{katsuragawa2019tip}
K.~Katsuragawa, J.~Wang, Z.~Shan, N.~Ouyang, O.~Abari, and D.~Vogel, ``Tip-tap: battery-free discrete 2d fingertip input,'' in \emph{Proceedings of the 32nd Annual ACM Symposium on User Interface Software and Technology}.\hskip 1em plus 0.5em minus 0.4em\relax New York, NY, USA: ACM, 2019, pp. 1045--1057.

\bibitem{ranganathan_rf_2018}
\BIBentryALTinterwordspacing
V.~Ranganathan, S.~Gupta, J.~Lester, J.~R. Smith, and D.~Tan, ``\BIBforeignlanguage{en}{{RF} {Bandaid}: {A} {Fully}-{Analog} and {Passive} {Wireless} {Interface} for {Wearable} {Sensors}},'' \emph{\BIBforeignlanguage{en}{Proceedings of the ACM on Interactive, Mobile, Wearable and Ubiquitous Technologies}}, vol.~2, no.~2, pp. 1--21, Jul. 2018. [Online]. Available: \url{https://dl.acm.org/doi/10.1145/3214282}
\BIBentrySTDinterwordspacing

\bibitem{gao2019livetag}
C.~Gao, Y.~Li, and X.~Zhang, ``Livetag: Sensing human-object interaction through passive chipless wi-fi tags,'' \emph{GetMobile: Mobile Computing and Communications}, vol.~22, no.~3, pp. 32--35, 2019.

\bibitem{Talla2017Sep}
V.~Talla, M.~Hessar, B.~Kellogg, A.~Najafi, J.~R. Smith, and S.~Gollakota, ``{LoRa Backscatter: Enabling The Vision of Ubiquitous Connectivity},'' \emph{Proc. ACM Interact. Mob. Wearable Ubiquitous Technol.}, vol.~1, no.~3, pp. 1--24, Sep. 2017.

\bibitem{fraternali_pible_2018}
\BIBentryALTinterwordspacing
F.~Fraternali, B.~Balaji, Y.~Agarwal, L.~Benini, and R.~Gupta, ``Pible: battery-free mote for perpetual indoor {BLE} applications,'' in \emph{Proceedings of the 5th {Conference} on {Systems} for {Built} {Environments}}, ser. {BuildSys} '18.\hskip 1em plus 0.5em minus 0.4em\relax New York, NY, USA: Association for Computing Machinery, Nov. 2018, pp. 168--171. [Online]. Available: \url{https://dl.acm.org/doi/10.1145/3276774.3282822}
\BIBentrySTDinterwordspacing

\bibitem{jeon_luxbeaconbatteryless_2019}
K.~E. Jeon, J.~She, J.~Xue, S.-H. Kim, and S.~Park, ``{luXbeacon}—{A} {Batteryless} {Beacon} for {Green} {IoT}: {Design}, {Modeling}, and {Field} {Tests},'' \emph{IEEE Internet of Things Journal}, vol.~6, no.~3, pp. 5001--5012, Jun. 2019, conference Name: IEEE Internet of Things Journal.

\bibitem{schmid_led_2013}
\BIBentryALTinterwordspacing
S.~Schmid, G.~Corbellini, S.~Mangold, and T.~R. Gross, ``\BIBforeignlanguage{en}{{LED}-to-{LED} visible light communication networks},'' in \emph{\BIBforeignlanguage{en}{Proceedings of the fourteenth {ACM} international symposium on {Mobile} ad hoc networking and computing}}.\hskip 1em plus 0.5em minus 0.4em\relax Bangalore India: ACM, Jul. 2013, pp. 1--10. [Online]. Available: \url{https://dl.acm.org/doi/10.1145/2491288.2491293}
\BIBentrySTDinterwordspacing

\bibitem{li_retro-vlc_2015}
\BIBentryALTinterwordspacing
J.~Li, A.~Liu, G.~Shen, L.~Li, C.~Sun, and F.~Zhao, ``\BIBforeignlanguage{en}{\textit{{Retro}-{VLC}}: {Enabling} {Battery}-free {Duplex} {Visible} {Light} {Communication} for {Mobile} and {IoT} {Applications}},'' in \emph{\BIBforeignlanguage{en}{Proceedings of the 16th {International} {Workshop} on {Mobile} {Computing} {Systems} and {Applications}}}.\hskip 1em plus 0.5em minus 0.4em\relax Santa Fe New Mexico USA: ACM, Feb. 2015, pp. 21--26. [Online]. Available: \url{https://dl.acm.org/doi/10.1145/2699343.2699354}
\BIBentrySTDinterwordspacing

\bibitem{xu_passivevlc_2017}
\BIBentryALTinterwordspacing
X.~Xu, Y.~Shen, J.~Yang, C.~Xu, G.~Shen, G.~Chen, and Y.~Ni, ``\BIBforeignlanguage{en}{{PassiveVLC}: {Enabling} {Practical} {Visible} {Light} {Backscatter} {Communication} for {Battery}-free {IoT} {Applications}},'' in \emph{\BIBforeignlanguage{en}{Proceedings of the 23rd {Annual} {International} {Conference} on {Mobile} {Computing} and {Networking}}}.\hskip 1em plus 0.5em minus 0.4em\relax Snowbird Utah USA: ACM, Oct. 2017, pp. 180--192. [Online]. Available: \url{https://dl.acm.org/doi/10.1145/3117811.3117843}
\BIBentrySTDinterwordspacing

\bibitem{wu_turboboosting_2020}
\BIBentryALTinterwordspacing
Y.~Wu, P.~Wang, K.~Xu, L.~Feng, and C.~Xu, ``\BIBforeignlanguage{en}{Turboboosting {Visible} {Light} {Backscatter} {Communication}},'' in \emph{\BIBforeignlanguage{en}{Proceedings of the {Annual} conference of the {ACM} {Special} {Interest} {Group} on {Data} {Communication} on the applications, technologies, architectures, and protocols for computer communication}}.\hskip 1em plus 0.5em minus 0.4em\relax Virtual Event USA: ACM, Jul. 2020, pp. 186--197. [Online]. Available: \url{https://dl.acm.org/doi/10.1145/3387514.3406229}
\BIBentrySTDinterwordspacing

\bibitem{xu_low-latency_2022}
\BIBentryALTinterwordspacing
K.~Xu, C.~Gong, B.~Liang, Y.~Wu, B.~Di, L.~Song, and C.~Xu, ``\BIBforeignlanguage{en}{Low-{Latency} {Visible} {Light} {Backscatter} {Networking} with {RetroMUMIMO}},'' in \emph{\BIBforeignlanguage{en}{Proceedings of the 20th {ACM} {Conference} on {Embedded} {Networked} {Sensor} {Systems}}}.\hskip 1em plus 0.5em minus 0.4em\relax Boston Massachusetts: ACM, Nov. 2022, pp. 448--461. [Online]. Available: \url{https://dl.acm.org/doi/10.1145/3560905.3568507}
\BIBentrySTDinterwordspacing

\bibitem{lightanchors2019uist}
\BIBentryALTinterwordspacing
K.~Ahuja, S.~Pareddy, R.~Xiao, M.~Goel, and C.~Harrison, ``Lightanchors: Appropriating point lights for spatially-anchored augmented reality interfaces,'' in \emph{Proceedings of the 32nd Annual ACM Symposium on User Interface Software and Technology}, ser. UIST '19.\hskip 1em plus 0.5em minus 0.4em\relax New York, NY, USA: Association for Computing Machinery, 2019, p. 189–196. [Online]. Available: \url{https://doi.org/10.1145/3332165.3347884}
\BIBentrySTDinterwordspacing

\bibitem{barman_every_2021}
\BIBentryALTinterwordspacing
L.~Barman, A.~Dumur, A.~Pyrgelis, and J.-P. Hubaux, ``\BIBforeignlanguage{en}{Every {Byte} {Matters}: {Traffic} {Analysis} of {Bluetooth} {Wearable} {Devices}},'' \emph{\BIBforeignlanguage{en}{Proceedings of the ACM on Interactive, Mobile, Wearable and Ubiquitous Technologies}}, vol.~5, no.~2, pp. 1--45, Jun. 2021. [Online]. Available: \url{https://dl.acm.org/doi/10.1145/3463512}
\BIBentrySTDinterwordspacing

\bibitem{do2023powering}
Y.~Do, N.~Arora, A.~Mirzazadeh, I.~Moon, E.~Xu, Z.~Zhang, G.~D. Abowd, and S.~Das, ``Powering for privacy: improving user trust in smart speaker microphones with intentional powering and perceptible assurance,'' in \emph{32nd USENIX Security Symposium (USENIX Security 23)}.\hskip 1em plus 0.5em minus 0.4em\relax Anaheim, CA, USA: USENIX, 2023, pp. 2473--2490.

\bibitem{de2025neuromorphic}
C.~De~Luca, M.~Tincani, G.~Indiveri, and E.~Donati, ``A neuromorphic multi-scale approach for real-time heart rate and state detection,'' \emph{npj Unconventional Computing}, vol.~2, no.~1, p.~6, 2025.

\bibitem{rivelli2025adaptively}
F.~Rivelli, M.~Popov, C.~S. Kouzinopoulos, and G.~Tang, ``Adaptively pruned spiking neural networks for energy-efficient intracortical neural decoding,'' \emph{arXiv preprint arXiv:2504.11568}, 2025.

\bibitem{wu2024lightweight}
Y.~Wu, F.~Paredes-Vall{\'e}s, and G.~C. De~Croon, ``Lightweight event-based optical flow estimation via iterative deblurring,'' in \emph{2024 IEEE International Conference on Robotics and Automation (ICRA)}.\hskip 1em plus 0.5em minus 0.4em\relax IEEE, 2024, pp. 14\,708--14\,715.

\bibitem{chen2024neuromorphic}
J.~Chen, S.~Park, P.~Popovski, H.~V. Poor, and O.~Simeone, ``Neuromorphic split computing with wake-up radios: Architecture and design via digital twinning,'' \emph{IEEE Transactions on Signal Processing}, 2024.

\bibitem{chakravarthi_recent_2024}
\BIBentryALTinterwordspacing
B.~Chakravarthi, A.~A. Verma, K.~Daniilidis, C.~Fermuller, and Y.~Yang, ``Recent {Event} {Camera} {Innovations}: {A} {Survey},'' Aug. 2024, arXiv:2408.13627 [cs]. [Online]. Available: \url{http://arxiv.org/abs/2408.13627}
\BIBentrySTDinterwordspacing

\bibitem{scott_neurocamtags_2024}
\BIBentryALTinterwordspacing
D.~Scott, M.~Bringle, I.~Fahad, G.~Morales, A.~Zahid, and S.~Swaminathan, ``{NeuroCamTags}: {Long}-{Range}, {Battery}-free, {Wireless} {Sensing} with {Neuromorphic} {Cameras},'' \emph{Proceedings of the ACM on Interactive, Mobile, Wearable and Ubiquitous Technologies}, vol.~8, no.~3, Sep. 2024. [Online]. Available: \url{https://dl.acm.org/doi/10.1145/3678529}
\BIBentrySTDinterwordspacing

\bibitem{psb:18:ten}
\BIBentryALTinterwordspacing
J.~S. Plank, C.~D. Schuman, G.~Bruer, M.~E. Dean, and G.~S. Rose, ``The {TENNLab} exploratory neuromorphic computing framework,'' \emph{IEEE Letters of the Computer Society}, vol.~1, no.~2, pp. 17--20, July-Dec 2018. [Online]. Available: \url{https://doi.ieeecomputersociety.org/10.1109/LOCS.2018.2885976}
\BIBentrySTDinterwordspacing

\bibitem{schuman_evolutionary_2020}
\BIBentryALTinterwordspacing
C.~D. Schuman, J.~P. Mitchell, R.~M. Patton, T.~E. Potok, and J.~S. Plank, ``Evolutionary {Optimization} for {Neuromorphic} {Systems},'' in \emph{Proceedings of the 2020 {Annual} {Neuro}-{Inspired} {Computational} {Elements} {Workshop}}, ser. {NICE} '20.\hskip 1em plus 0.5em minus 0.4em\relax New York, NY, USA: Association for Computing Machinery, Jun. 2020, pp. 1--9. [Online]. Available: \url{https://dl.acm.org/doi/10.1145/3381755.3381758}
\BIBentrySTDinterwordspacing

\bibitem{spb:19:nte}
C.~D. Schuman, J.~S. Plank, G.~Bruer, and J.~Anantharaj, ``Non-traditional input encoding schemes for spiking neuromorphic systems,'' in \emph{IJCNN: The International Joint Conference on Neural Networks}, Budapest, 2019, pp. 1--10.

\bibitem{prw:25:cpa}
\BIBentryALTinterwordspacing
J.~S. Plank, C.~P. Rizzo, C.~A. White, and C.~D. Schuman, ``The cart-pole application as a benchmark for neuromorphic computing,'' \emph{Journal of Low Power Electronics and Applications}, vol.~15, no.~1, pp. 1--27, 2025. [Online]. Available: \url{https://www.mdpi.com/2079-9268/15/1/5}
\BIBentrySTDinterwordspacing

\bibitem{pdg:24:risp}
J.~S. Plank, K.~E.~M. Dent, B.~Gullett, C.~P. Rizzo, and C.~D. Schuman, ``The {RISP} neuroprocessor -- open source support for embedded neuromorphic computing,'' in \emph{IEEE International Conference on Rebooting Computing (ICRC)}, San Diego, December 2024.

\end{thebibliography}

\end{document}